# A mini review on NiFe-based materials as highly active oxygen evolution reaction electrocatalysts

Ming Gong[1] and Hongjie Dai[1]*

[1]Department of Chemistry, Stanford University, Stanford, CA 94305, USA

*Correspondence to: hdai@stanford.edu

**Abstract**

Oxygen evolution reaction (OER) electrolysis, as an important reaction involved in water splitting and rechargeable metal-air battery, has attracted increasing attention for clean energy generation and efficient energy storage. Nickel/iron (NiFe)-based compounds have been known as active OER catalysts since the last century, and renewed interest has been witnessed in recent years on developing advanced NiFe-based materials for better activity and stability. In this review, we present the early discovery and recent progress on NiFe-based OER electrocatalysts in terms of chemical properties, synthetic methodologies and catalytic performances. The advantages and disadvantages of each class of NiFe-based compounds are summarized, including NiFe alloys, electro-deposited films and layered-double hydroxide nanoplates. Some mechanistic studies of the active phase of NiFe-based compounds are introduced and discussed to give insight into the nature of active catalytic site, which could facilitate further improving NiFe based OER electrocatalysts. Finally, some applications of NiFe-based compounds for OER are described, including the development of electrolyzer operating with a single AAA battery with voltage below 1.5V and high performance rechargeable Zn-air batteries.

# Introduction

The increasing demand for energy and diminishing natural energy resources have led to a revolutionary era of discovering earth-abundant energy alternatives and designing efficient energy-storage devices[1-7]. Hydrogen ($H_2$), with its high mass-specific energy density, has been considered to be a promising energy source and a substitute for fossil fuels[8-10]. Currently $H_2$ production still mostly relies on the industry of fossil fuels and suffers from low purity and high cost[8]. One of the most efficient ways of producing $H_2$ at low cost and high purity is water splitting into hydrogen and oxygen by electricity or sunlight[2, 3, 6, 11-16]. Oxygen evolution reaction (OER), as an important half-reaction involved in water splitting, has been intensely investigated for decades[11, 17-20]. OER is a demanding step that includes four proton-coupled electron transfers and oxygen-oxygen bond formation, so it is kinetically not favored and requires catalyst for expediting the reaction[21, 22]. Iridium dioxide ($IrO_2$) and ruthenium dioxide ($RuO_2$) are the state-of-the-art OER electrocatalysts with low overpotential and Tafel slope, especially in acidic conditions[6, 23-25]. Nevertheless, these catalysts suffer from scarcity and high cost, and cannot be utilized in water splitting industries to obtain economic $H_2$ energy resources.

Nickel (Ni) is an earth-abundant first row transitional metal with corrosion-resistance and ductility. Ni and its oxide were discovered to exhibit electro-catalytic activity towards OER in alkaline solution in the early last century[26, 27]. One of the most interesting discoveries was that as nickel was always found in combination with iron (Fe) on earth and Fe impurities in the nickel hydroxide ($Ni(OH)_2$) electrodes can cause detrimental effects on Ni-based alkaline batteries by greatly lowering the OER overpotential[28-32]. The discovery had inspired many scientists to study the phenomenon, optimizing Fe content and synthesizing various NiFe mixed compounds for obtaining better OER electrocatalyts[33-35]. Up to now, the catalyst utilized in water electrolysis industry is stainless steel, which is often an alloy consisting of nickel, iron and chromium[13, 36-38]. Recently, with the rising demand for clean and renewable energy, researchers have revisited the field and designed more advanced NiFe-based OER electrocatalysts with higher catalytic activity for efficient and durable water electrolysis.

In this review, we will summarize the discovery and recent progress of NiFe-based compounds for OER electrocatalysis in terms of synthesis, catalytic performance, application and mechanism. We will start with a brief introduction about OER electrocatalysis followed by a summary of early discovery and optimization strategy of NiFe-based OER electrocatalysts. Then, recent progress on some advanced NiFe-based materials will be reviewed according to their chemical structures, synthetic methods and catalytic activities. Further, some insight into the mechanism of OER electrocatalysis by NiFe compounds and possible future direction to improve the performance NiFe OER electrocatalysts will be discussed. Finally, the applications of NiFe-based OER catalysts including water splitting and rechargeable metal air batteries will be briefly introduced. Through this review, we hope to provide the readers with a distinct perspective of the history, present and future of this field.

**OER electrocatalysis**

Water oxidation or OER is a reaction of generating molecular oxygen through several proton/electron-coupled processes[39, 40]. In acidic conditions, the reaction operates through oxidation of two water molecules ($H_2O$) into four protons ($H^+$) and one oxygen molecule ($O_2$) by losing a total of four electrons[40]. In basic conditions, oxidation of hydroxyl groups ($OH^-$) takes the lead and transforms into $H_2O$ molecules and $O_2$ molecule with the same four electrons involved[39].

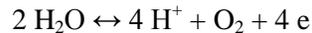

$$2\ H_2O \leftrightarrow 4\ H^+ + O_2 + 4\ e$$

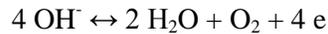

$$4\ OH^- \leftrightarrow 2\ H_2O + O_2 + 4\ e$$

As OER is an electron-coupled uphill reaction, electricity is usually used as energy input to drive the reaction. The standard potential for OER at pH 0 (as shown in equation 1 when the concentration of $H^+$ is 1 M) is 1.23 V vs normal hydrogen electrode (NHE). As the reaction involves $H^+$ or $OH^-$, the potential is dependent on pH by shifting 59 mV for each pH unit increase according to Nernst equation.

$$E = E^o - \frac{RT}{nF}\ln\frac{[Red]}{[Ox]}$$

($E$ is the cell potential, $E^o$ is the cell potential at standard conditions, $R$ is the ideal gas constant, $T$ is the temperature in kelvin, $n$ is the mole of electrons involved each mole of reaction, $F$ is the Faradaic's constant, [$Red$] is the concentration of reduced molecules and [$Ox$] is the concentration of oxidized molecules)

To remove the pH impact on the potential applied, people have introduced a reversible hydrogen electrode (RHE) reference by taking into account the pH shift, thus the theoretical potential required for OER is always 1.23 V vs RHE at all pHs. As shown in equation 1 and 2, generation of each $O_2$ molecule requires a transfer of four electrons, and as multiple electron transfer at one time is not kinetically favorable, OER usually involves multiple steps with one electron transfer per step[39]. An accumulation of the energy barrier in each step leads to the sluggish kinetics of OER with large overpotential to overcome. Therefore OER electrocatalysis is desired to expedite the reaction and lower the potential. As very few non-precious metal oxides can survive under oxidative potential in acidic condition[24], researchers have been searching for non-precious metal-based candidates for OER electrocatalysis in alkaline conditions, in which most metal oxides or hydroxides are chemically stable. An ideal OER electrocatalyst should have low overpotential, high durability, low-cost, earth-abundance and scalability. However, up to now catalysts meeting these requirements have not yet been discovered. NiFe-based compound is one of the most promising candidates with the highest OER electro-catalytic activities among all non-precious metal-based electrocatalysts.

**Early discovery**

The poisoning effect of Fe impurities on the $Ni(OH)_2$ electrodes in alkaline batteries were first found by Edison and Junger soon after their discovery of Ni-based alkaline batteries. Systematic study on the role of iron contamination revealed that increasing the iron contamination degree led to decreasing capacity and cycle life of the alkaline batteries by lowering the OER overpotential[28-32]. Such effect could be observed with as low concentration as <1 % Fe and attributed to the catalytic behavior of the conducting ferric oxide in the study by Mlynarek et al.[35]. With the rising interest in water splitting in the 1980s, Corrigan et al. first studied the oxygen evolution behavior of nickel oxide electrodes with Fe impurities. Fe impurities from the electrolyte co-precipitated onto the nickel oxide films were discovered to share similarly strong effects on catalyzing oxygen evolution reactions[34]. The decrease in OER overpotential and decrease in discharge capacity was reported even at an ultra-low Fe concentration (0.01%) (Figure 1), demonstrating the high sensitivity nature of OER on Ni-based electrodes to Fe impurities. By adjusting Fe content, a composite NiFe hydrous oxide with >10% Fe showed intriguing activities toward OER electrocatalysis with ~200-250 mV overpotential at a current density of 10 mA/cm$^2$ and a low Tafel slope of 20-25 mV/decade (Figure 1b). In addition, Corrigan et al. also introduced other metal ions to $Ni(OH)_2$ films by electro-deposition and the film with co-deposited $Fe/Ni(OH)_2$ exhibited high OER electro-catalytic activity to afford a 287 mV overpotential at 16 mA/cm$^2$, outperforming all other transitional metal element introduced except cerium. These results suggested a promising future of NiFe-based compound for low-cost OER electrocatalysis[41].

**NiFe alloy**

The affinity of Ni (element No. 28) and Fe (element No.26) facilitates the formation of a NiFe alloy, as indicated by steel and abundance of NiFe alloy in the earth's core. One of the most facile methods of synthesizing NiFe-based electrocatalysts is to mechanically mix metallic Ni and Fe into NiFe alloy. For example, amorphous Ni-Fe-Si-B alloys prepared by melt spinning on a CuZr drum showed an OER overpotential of ~700 mV vs mercury/mercury oxide reference electrode (Hg/HgO) (~340 mV in overpotential) at 100 mA/cm$^2$ in 30 wt% potassium hydroxide (KOH) under room temperature[42]. A later work demonstrated the synthesis of various combinations of cobalt, nickel, molybdenum and iron alloys by mechanical alloying. The alloys showed improved OER electrocatalytic activity with 11-234 mA/cm$^2$ at an overpotential of 400 mV under different temperatures of 298 K, 323 K and 343 K[43]. The mechanical alloying process is a good approach of mixing a wide range of elements to study the synergistic effects on different elements, but the physical mixture nature leads to large crystal size with low electrochemically accessible surface area and lack of chemical contact between different elements. The incapability of producing novel chemical structures by physical alloying limits its potential in exploring novel structures for OER electrocatalysis.

An alternative way of synthesizing NiFe alloy is by cathodic electro-reduction of mixed metal salt solutions. Since nanoscale NiFe alloy often turns into NiFe oxide when exposed to air especially at the surface, people sometimes reported the catalyst as electrodeposited NiFe oxide. Such approach could afford optimal adherence of the catalytic active material to the underlying substrate, thereby lowering the overpotential introduced by ohmic loss. By tuning the electrolyte composition and electro-deposition parameters such as current and time, NiFe alloys with a variety of morphologies and catalytic properties could be synthesized. Potvin et al. first reported a citrate solution with metal sulfate salt as electrolyte and derived a particle-based film of NiFe alloy early in 1992[44]. Among the films with different Ni/Fe ratios, the 45/55 at.% Ni-Fe deposits outperformed with a current density of 351 mA/cm$^2$ at an overpotential of 350 mV in 40 wt. % KOH at 80 $^o$C. Both the cathodic electro-deposition condition and pre-anodization condition were found to greatly influence the OER catalytic activity, suggesting the possibility of improving catalyst performance by fine-tuning the electro-synthesis condition. Ascorbic acid[45], boric acid[46-50] and ammonium sulphate[24, 51-54] solutions containing iron and nickel sulphate were later discovered as suitable electrolytes for cathodic electro deposition of NiFe films, and recently the latter two electrolytes have been widely used for NiFe alloy electro-synthesis. Grande et al. studied the NiFe thin films electrodeposited in boric acid containing electrolyte, and addition of boric acid solutions was found to be essential for NiFe co-deposition by decreasing the rate of Ni deposition while increasing the rate of Fe deposition[46]. As appropriate Fe content was earlier demonstrated to be a significant factor impacting the OER electrocatalytic activity, increasing Fe content by the boric acid approach could potentially improve the activity as confirmed by later studies finding excellent electrocatalytic activity with uniform NiFe films deposited in a boric acid solution[47-50]. Merrill et al. further explored an ammonium sulphate solution for NiFe electro-deposition and they compared the catalyst with other mixed transition metal deposited in similar conditions[54]. The as-synthesized NiFe film outperformed other catalysts by at least 50 mV with an incredibly low Tafel slope of 14.8 mV/decade, corroborating the highly active nature of NiFe-based materials for OER catalysis. By adopting similar synthetic approach, Louie et al. optimized Ni/Fe ratio and the NiFe catalyst containing 40% Fe exhibited the highest activity among the catalyst[53]. In situ Raman spectroscopy of the deposited NiFe film elucidated the correlation of the average Ni oxidation state with the OER catalytic activity, which will be discussed later in the review.

In addition, the synthetic method of electro-deposition in ammonium sulphate was further utilized in McCrory et al.'s study of evaluating the activities of a series of heterogeneous electrocatalyst benchmarks[24]. In concentrated alkaline condition of 1 M sodium hydroxide (NaOH), NiFe-based compound stood out among all non-precious metal catalysts with nearly 10 times higher activity per electrochemically active surface area and stable catalytic behavior over 2 hours, suggesting useful NiFe-based electrocatalysts for OER catalysis applications (Figure 2). The ease of NiFe alloy synthesis by electro deposition and the excellent adherence to the substrate makes it a popular approach for studying NiFe-based compound. However, the limitation of NiFe alloy lies in that structural transformation is needed before entering an OER active phase. Such structural evolution often confines to the surface where the electrolyte is accessible, implying the necessity of NiFe deposits with high surface area for further improving the catalyst. Even though some recent studies demonstrated the electro-deposition of dendritic NiFe alloys with high OER performance[55], few reports have been dedicated to highly porous NiFe films with large surface area. In addition, it still remained unclear whether the oxidation process of NiFe film fully penetrated through the entire film or only on the surface, posing a challenge to structural analysis both on the surface and beneath the surface in determining the active phase for OER electrocatalysis.

**NiFe oxide**

Compared to the metallic state of Ni and Fe element in NiFe alloy, NiFe oxide often has oxidation states of +2 and +3 for Ni and Fe respectively, which are much closer to the oxidation states in OER region. To derive the NiFe oxide phase, annealing at high temperature is usually required and therefore most NiFe oxides are crystalline with well defined structure[56-63] (often with X-ray diffraction analysis). The preparation of NiFe oxide could be afforded by a variety of synthetic methods. The earliest approach was by reactive sputtering co-deposition with a NiFe alloy target and a 20% oxygen/argon atmosphere with 10 mTorr pressure[64]. Similar discovery of tremendously improved activity by Fe incorporation into Ni oxide films was obtained, and the best sputtered sample exhibited a current density of 80 mA/cm$^2$ at an overpotential of 362 mV with a Tafel slope of less than 40 mV/decade. Importantly, due to the crystalline nature of the sputtered NiFe oxide, an impressive durability of over 7000 hours was observed[64].

With a Ni/Fe ratio of ½, the NiFe oxide can form nickel ferrite (NiFe$_2$O$_4$) with a spinel structure, showing cubic closed-packed oxide anions with Ni$^{2+}$ occupying one-eighth of the tetrahedral holes and Fe$^{3+}$ occupying half of the octahedral holes. Owing to the highly corrosion-resistant nature (insoluble in most acids), nickel ferrite has attracted much attention as an OER electrocatalyst, even though it is more often investigated as a magnetic material due to the strong magnetism[65, 66]. The nickel ferrite catalyst could be produced through low-temperature precipitation followed by high temperature annealing. Some earlier studies shared similar conclusion that NiFe mixed compound (spinel NiFe$_2$O$_4$) outperforms Fe alone (spinel Fe$_3$O$_4$) and Ni alone (cubic NiO) in OER electro-catalytic activity[61]. Due to the heterogeneity of spinel structures, researchers have recently focused on substituting the NiFe$_2$O$_4$ by other cations (e.g. Cr, V, Mo and Co) for enhancing the electro-catalytic activity[56-59]. Even though improved OER activity were observed, the NiFe$_2$O$_4$ or other crystalline NiFe oxide catalysts seemed to underperform the electro-deposited NiFe films probably due to the rigidity of the crystalline NiFe oxides.

Though most NiFe oxide are crystalline resulted from high temperature annealing, one exception is the production of amorphous NiFe oxide by thermal decomposition of metal organic compound at relatively lower temperatures. Smith et al. introduced a facile photochemical metal-organic deposition (PMOD) method for preparing amorphous (mixed) metal oxide thin films with thickness of 150 to 200 nm (Figure 3a) [67]. Interestingly, amorphous films synthesized at 100 $^{o}$C was found to have better OER electro-catalytic performance than crystalline films formed at 600 $^{o}$C (Figure 3d). By comparing various oxide films synthesized by the same PMOD method, mixed NiFe oxide and mixed oxide containing Ni and Fe were confirmed to outperform other catalysts with low overpotential and low Tafel slope (Figure 3b-3d). However, the catalyst tended to decay slowly over time probably due to the amorphous nature. Therefore, the balance between crystallinity to obtain high durability with rigid structures under OER operation and amorphousness to obtain high activity with large electrochemically active surface area (ECSA) is essential for developing advanced NiFe-based OER electrocatalyst.

**NiFe layered double hydroxide**

Layered double hydroxide (LDH) is a class of layered materials consisting of positively charged layers and charge-balancing anions in the interlayer region[68-70]. The positively charged layers are constructed by partially substituting divalent cations (e.g. $Ni^{2+}$, $Mg^{2+}$, $Ca^{2+}$, $Mn^{2+}$, $Co^{2+}$, $Cu^{2+}$, $Zn^{2+}$) or monovalent cations (e.g. $Li^{+}$) by trivalent cations (e.g. $Al^{3+}$, $Co^{3+}$, $Fe^{3+}$, $Cr^{3+}$). The intercalated anions are typical carbonate ($CO_3^{2-}$), but it could be easily replaced by other anions (e.g. $NO_3^{-}$, $SO_4^{2-}$, $Cl^{-}$, $Br^{-}$). Therefore, the LDH structure is attractive to electrochemical process as it is highly accessible to electrolyte by anion exchange. Synthesis of NiFe LDH was achieved by earlier studies[71, 72], but NiFe LDH for oxygen evolution electrocatalysis has been barely studied until recently. One of the major issues associated with NiFe LDH is its low conductivity, a common problem with hydroxides[73]. Researchers have attempted to solve the problem by either hybridizing with conductive material or directly growing LDH on conductive substrate[73-76].

Our group was the first to develop a NiFe LDH with or without carbon nanotubes as support for OER electrocatalysis (Figure 4a)[73]. By slow hydrolysis of metal salt at low temperature followed by a high-temperature solvothermal step for crystallization, we obtained ultrathin NiFe LDH nanoplate (~ 5 nm) covalently attached to a multi-walled carbon nanotube (Figure 4b). The hybrid material could deliver a current density of 5 mA/cm$^2$ at a overpotential of 250mV with a Tafel slope of 31 mV/decade in 1 M KOH at a loading of 0.25 mg/cm$^2$ (Figure 4c). The LDH nanoplate alone without nanotubes showed a slightly lower OER catalytic activity by ~30 mV on substrates due to relatively low conductivity, but if loaded on high conductive Ni foam, NiFe LDH nanoplate could afford high activity as well as high stability for at least 4 days (Figure 4d). Therefore, the high performance of the material was mainly attributed to the LDH phase and further facilitated by the ultrathin nature of nanoplates and strong coupling to carbon nanotubes.

The high intrinsic OER electro-catalytic activity of NiFe LDH was further confirmed by later studies of NiFe LDH grown on graphene[74], carbon quantum dot[75] and Ni foam substrate[76]. To glean into the active site of NiFe LDH, Trotochaud et al. systematically studied the effect of Fe incorporation into Ni(OH)$_2$ film (NiFe LDH structure) on OER catalytic activity[77]. Surprisingly, a >30-fold increase in conductivity was observed upon Fe incorporation into the Ni(OH)$_2$ film. However, since the recording i-V curves were iR-compensated, the conductivity improvement could not explain the OER current boost. Further study of film thickness effect on turnover frequency (TOF) showed a lower dependence of thickness for co-deposited Fe films than Fe-free films, suggesting an Fe-induced partial-charge-mechanism. Another important conclusion of the study was that upon aging, Ni(OH)$_2$ could increase its crystallinity as well as collect Fe impurities in the electrolyte[77], illustrating the cause of improved activity by aged Ni(OH)$_2$ film and the high electro-catalytic activity of NiFe LDH phase.

Due to the large interlayer spacing between individual layers in the LDH structure, exfoliation of LDH structure into few layers and single sheets like has been demonstrated to expose more surface sites[69]. Song et al. demonstrated a synthetic route of preparing exfoliated LDH by an anion exchange step to expand interlayer spacing and a following delamination step in formamide solution (Figure 5a) [78]. The as-synthesized exfoliated LDH nanosheet demonstrated an average thickness of ~0.8 nm, implying single or double layered structure (Figure 5c). Upon exfoliation, significantly higher OER electro-catalytic activity was observed with a variety of LDHs including NiFe LDH (Figure 5b). The authors proposed that a large number of active sites were exposed after exfoliation, as they observed dramatically improved OER activity with similar ECSAs (Figure 5d). However, further spectroscopy and theoretical calculation studies were needed to confirm the active site of NiFe LDH.

NiFe LDH has recently been demonstrated to possess excellent electro-catalytic activity and durability for OER catalysis; however, the active site of NiFe LDH remained unclear, and further mechanistic studies could help to understand the structural relevance to OER electro-catalytic activity and further improve NiFe LDH-based electrocatalyst by constructing more advanced structures to tune the electronic property.

**Screening NiFe-based catalysts containing other elements**

One straightforward approach of developing NiFe-based electrocatalysts with better activities is to introduce other elements into the structures. However, changing different elements and tuning element ratios demand tremendous efforts. Therefore, researchers have designed high-throughput screening methods by $O_2$ probes to visualize the OER electro-catalytic activity trend while tuning the elements[79-81]. Consequently, more OER catalyst candidates could be screened during a single scan. Gerken et al. utilized fluorescent pressure-sensitive paints that can vary its intensity upon detecting different partial pressures of $O_2$[79]. Improved quantitative measurements were enabled by two fluorophores with one insensitive to $O_2$ and the other quenching fluorescence in proportion to partial pressure of $O_2$ (Figure 6a). Twenty-one different ternary metal-oxide combinations were screened and each ternary-oxide measurement contained twenty-one samples with different element ratio. The intensity of each individual point was eventually plotted into a triad showing different OER catalytic activity with different elemental ratios for each ternary-oxide measurement. Through comparing different triads, a ternary oxide composing of a combination of 60% Ni / 20 % Al / 20 % Fe was demonstrated to exhibit the highest catalytic activity, which was consistent with the high intrinsic activity of NiFe oxide over OER catalysis (Figure 6b). The high OER catalytic activity of ternary NiAlFe compound was corroborated by their later studies on inverse spinel $NiFeAlO_4$ structures[82]. The team further improved the throughput by an order-of-magnitude with 15*14 rectangular arrays of varied oxide compositions. By screening 3500 earth abundant mixed oxide catalysts, NiFe oxides containing a third metal (e.g. Al, Ga and Cr) were confirmed to be most active OER catalysts in alkaline conditions[80]. The triads containing ternary oxide were further extended to a quaternary (Ni-Fe-Co-Ce) oxide system, in which 5456 different compositions were investigated by a similar high-throughput screening method using $O_2$-sensitive probes[81]. An interesting pseudo-ternary cross-section containing 665 different compositions was discovered to possess better catalytic performance over other combinations.

The high throughput screening method provides an efficient and cost-effective way of discovering new OER-catalytic active material with a variety of element combinations. However, currently only materials with simple synthetic methods can be screened, but advanced OER catalysts might need complicated synthesis at harsh conditions. Therefore, the most efficient way of discovering and designing novel OER catalysts with better performance might be using modern synthetic methodology under the guidance of theoretical calculation and high throughput screens.

**Mechanism**

Even though a variety of NiFe-based catalysts have been developed, the active phase and detailed mechanism of NiFe-based compounds still remain unclear and partly in debate. The controversy stems from the difficulty of identifying the structure and pathway both experimentally and theoretically during OER electrocatalysis. The earliest related mechanistic studies researched into the active phase of $Ni(OH)_2$ or nickel oxide ($NiO_x$) catalyzing OER. The phase transformation of $Ni(OH)_2$-based electrode followed the Bode's diagram with β-$Ni(OH)_2$ and α-$Ni(OH)_2$ changing into β-nickel oxyhydroxide (β-NiOOH) and γ-NiOOH respectively during charging and discharging. γ-NiOOH can also be derived from overcharging β-NiOOH, while α-$Ni(OH)_2$ turns into β-$Ni(OH)_2$ over aging[83]. The α-$Ni(OH)_2$/γ-NiOOH pathway generally involves more electron transfer and structural change than the β-$Ni(OH)_2$/ β-NiOOH pathway[84]. Over a long period of time, β-NiOOH has been generally considered as the active phase for OER catalysis due to the observation of improved OER catalytic activity after aging[83, 85, 86]. However, recent studies leaned towards an active phase of γ-NiOOH. By using in situ X-ray absorption near-edge structure (XANES) spectroscopy, a NiOOH thin film deposited form nickel-containing borate electrolyte was first found to exhibit an average oxidation state of +3.6, indicating a predominant portion of γ-NiOOH during OER electrocatalysis[87]. Further, by studying the impact of Fe impurity upon the OER activity of NiOOH film, Trotochaud et al. argued that during aging process the NiOOH film could gradually collect the Fe impurities remained in the solution with modified electronic properties, which could greatly contribute to the improved activity, thereby challenging the long-held view of a more active β-NiOOH phase[77].

Study on the effect of Fe incorporation on Ni-based OER electrocatalysis was further accessed by a number of techniques [e.g density functional theory (DFT) calculation and in situ spectroscopy]. Corrigan et al. first attributed the activity gain to the improvement in electrical conductivity by Fe incorporation[34]. The better conductivity was later confirmed by steady-state in situ film conductivity measurement, but the huge overpotential difference at low current densities could not be explained by a more conductive Fe-incorporated film[77]. Also, the iR-compensation usually adopted in most of the electrocatalysis studies could minimize the conductivity impact on OER catalytic performance.

In situ spectroscopies have recently attracted increasing attention as useful tools for probing the catalyst while operated in catalytic conditions. In situ extended X-ray absorption fine structure (EXAFS) revealed that under oxygen evolution reaction conditions the average coordination number of Fe atoms increases, corroborating the participation of Fe in enhancing OER catalytic activity[88]. According to the observed crystalline $NiFe_2O_4$ phases by X-ray diffraction (XRD) as well as the high coordination number of octahedrally coordinated Fe in $NiFe_2O_4$, the active phase of $NiFe_2O_4$ was postulated. The possibly of other NiFe-based compounds formed by in situ electrochemical oxidation (especially at the surface) as contributing phases to the activity could not be fully ruled out due to the mostly amorphous nature of electrochemical synthesized materials.

The active phase of $NiFe_2O_4$ was later challenged by Louie et al. using in situ Raman spectroscopy[53]. Raman spectra of the NiFe oxide films under bias exhibited no traces of $NiFe_2O_4$ with typical bands at 700 cm$^{-1}$, implying that no $NiFe_2O_4$ was formed at the surface under oxidative potentials (Figure 7a). The polarized NiFe oxide films showed typical Raman peaks of NiOOH at 475 cm$^{-1}$ and 555 cm$^{-1}$ even at Fe contents of 90% or higher, indicating a possible active phase of NiOOH for OER electrocatalysis even with the presence of Fe (Figure 7b). Additionally, the Ni-O environment in NiFe oxide without bias was dramatically modified by Fe incorporation, especially at higher Fe content. Broader peaks with higher degrees of disorder were observed at high Fe/Ni ratios (Figure 7b). Such structural modification by Fe introduction might give rise to the improved OER activity.

DFT calculation has been widely employed as a prediction or supporting tool for various electrochemical processes, but few theoretical studies have been focusing on Ni-based or NiFe-based materials for OER electrocatalysis. Recently, Li et al. performed spin-polarized DFT calculations on different combinations of Ni and NiFe oxides[89]. An activity trend of Fe-doped β-NiOOH (0.26 V) > $NiFe_2O_4$ > β-NiOOH > Fe-doped γ-NiOOH > γ-NiOOH > $Fe_3O_4$ was obtained by calculating the overpotential required for OER. The Fe-doped β-NiOOH even possessed a theoretically better activity than $RuO_2$, a well-known OER catalyst. Nevertheless, one limitation of this study was that the activity trend was calculated using the elemental steps of OER in acidic conditions involving four proton-coupled electron transfers. The NiFe-based materials were demonstrated as highly active catalysts in basic conditions, which proceeded through electron transfers from hydroxyl groups to water molecules with the release of oxygen gas. Therefore, the different pathway could lead to the different intermediate and the different activity trend. However, the difficulty of DFT calculation in basic conditions lies in the charged intermediate during each elemental step, which demands tremendous time and effort[12].

Overall, various attempts to glean the mechanism of improved activity with NiFe-based compounds have been conducted by spectroscopic studies and theoretical calculations. A convincing and congruous theory has yet to emerge due to structural limitations of NiFe compounds generated in situ on the surface and calculation limitations of introducing charged intermediate. The mechanistic studies were still desired to shine light on the understanding of OER processes and develop more OER electrocatalysts with better activity.

**Applications**

One of the most eminent applications for OER is water electrolysis, which is a promising way of producing high-purity hydrogen gas with low cost for hydrogen fuel cell powered electric vehicles with zero emission. Our group recently paired up a NiFe LDH catalyst serving as an oxygen-generating anode and a nickel oxide/nickel hetero-structured catalyst on CNTs (NiO/Ni-CNT) serving as a hydrogen generating cathode in 1 M KOH to make an alkaline electrolyzer (Figure 8a)[12]. The device could achieve a current density of 20 mA/cm$^2$ at a voltage as low as 1.5 V and a current density of 100 mA/cm$^2$ at a voltage of 1.58 V with both electrodes under loadings of 8 mg/cm$^2$ at room temperature (Figure 8b and 8c). Further increasing the operation temperature to 60$^\circ$C could greatly benefit the reaction kinetics and lower the voltage to 1.4 V (Figure 8b and 8c). The low voltage required can enable the device to be powered by a single alkaline AAA battery with gas bubbles evolving on both electrodes (Figure 8d). Notably, this was the first time to achieve < 1.5 V voltage for electrolyzers with non-precious metal electrocatalysts. With the low-cost, high activity and facile synthesis, the NiFe-based and NiO/Ni-based electrocatalysts have great potential for future electrolyzers. Improving the durability of the electrocatalysts could further push their industrial applications.

OER electrocatalyst could also be used in rechargeable metal-air battery as the positive electrode for charging the battery. During charging process, water is oxidized on the positive electrode and metal is deposited (reduced) on the negative electrode, while during discharging process oxygen is reduced on the positive electrode and metal is dissolved (oxidized) on the negative electrode. One of the most promising rechargeable metal-air batteries is the zinc-air battery due to its high volumetric energy density along with environmental friendliness, low cost and high safety[90]. The rechargeable zinc-air battery usually operates in concentrated alkaline solution with dissolved zincate, which matches the working region of NiFe-based material for OER electrocatalysis. Therefore, we integrated our NiFe LDH electrocatalyst into a primary zinc-air battery consisting of an oxygen reduction reaction (ORR) electrocatalyst with cobalt oxide/nitrogen-doped carbon nanotube (CoO/NCNT)[91] as the cathode and zinc foil as the anode to make a tri-electrode rechargeable zinc-air battery[92] (Figure 9a). The battery showed an unprecedented small charge-discharge polarization voltage of ~0.70 V at 20 mA/cm$^2$, matching the performance of precious metal-based catalysts with Pt/C as ORR catalyst and Ir/C as OER catalyst (Figure 9b). Importantly, the zinc-air battery with NiFe LDH and CoO/NCNT exhibited negligible voltage change or material degradation during both charging and discharging (Figure 9c and 9d), while the one with Pt/C and Ir/C showed drastic decay over time on the Ir/C side (Figure 9c). Therefore, NiFe-based materials, as promising OER electrocatalysts, can be utilized in advanced energy storage applications such as rechargeable zinc-air batteries.

**Conclusions**

This review highlights the old discovery and recent progress of NiFe-based compounds for oxygen evolution electrocatalysis. OER is a kinetically sluggish reaction that requires electrocatalysts to expedite the reaction, and NiFe-based materials are promising candidates with low cost and high activity in alkaline conditions. Various NiFe compounds with distinct chemical structures, physical morphologies and catalytic properties have been synthesized and investigated. Eletro-deposited NiFe alloy film, one of the earliest studied materials, could afford optimal electrical contact with the underlying substrate but show moderate activity and low durability due to the necessary phase transformation prior to OER catalysis. High-temperature annealed NiFe oxide exhibited high durability but relatively low activity due to the highly crystalline phase and chemical stability of mixed NiFe oxide (especially for corrosion-resistant $NiFe_2O_4$). NiFe LDH was a rising candidate with the highest activity owing to large electrochemically active surface area but it still suffered from the low electrical conductivity. High through-put screening analysis was also developed for selecting advanced NiFe-based materials containing other elements with better performance. Even though tremendous efforts and progress have been made to develop novel NiFe-based material with better activity and stability, few reports have elaborated the detailed mechanism of improved activity by Fe incorporation. Existing mechanistic studies confirmed the modification of structural and electronic properties with Fe introduction, and a NiOOH-based active phase was corroborated by in situ spectroscopy technologies. The cause of greatly enhanced OER catalysis by Fe still remained unclear, and should be emphasized by future studies. Possible approaches might be in situ spectroscopy and theoretical calculation with applied potential. Understanding the basic mechanism could greatly facilitate the discovery of advanced NiFe-based OER electrocatalysts with higher activity and stability, which would be essential in energy conversion and energy storage applications such as water splitting and rechargeable metal-air batteries.

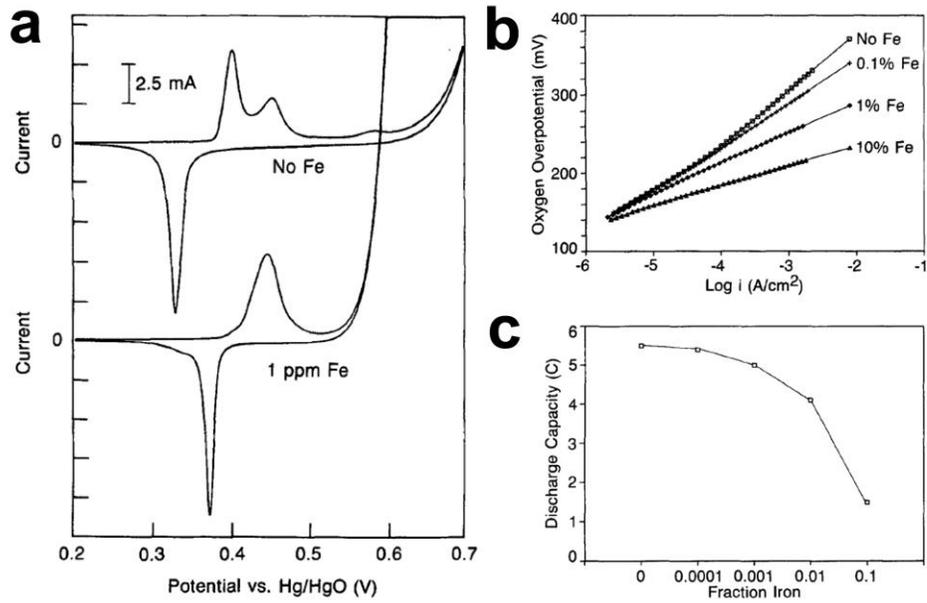

Figure 1. a) Iron impurity effect in the electrolyte on nickel electrode by cyclic voltammetry at a scan rate 10 mV/s. b) Tafel plots of oxygen evolution at thin nickel oxide film electrodes with different amount of coprecipitated iron by open-circuit decay measurements. c) Effect of coprecipitated iron on the discharging capacity of at thin nickel oxide film electrodes at a discharge current density of 8 mA/cm$^2$. (Reprinted with permission from Ref. 34.. Copyright 1987, The Electrochemical Society)

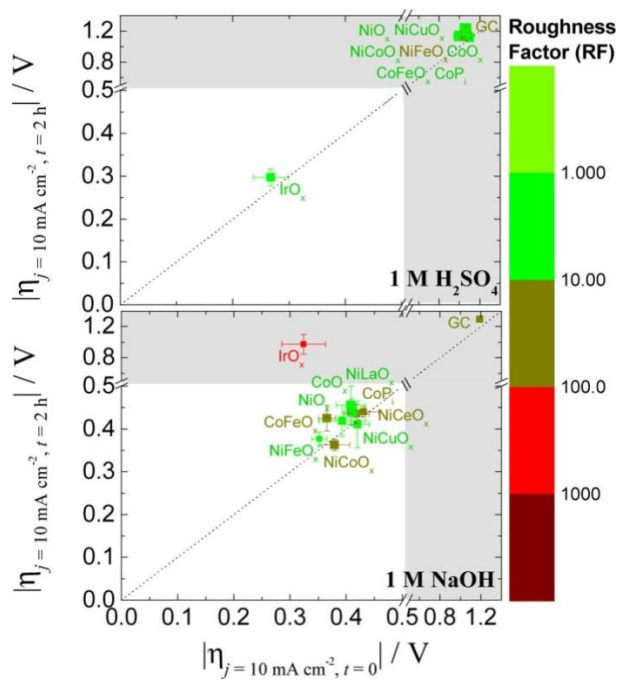

Figure 2. Comprehensive plots of catalytic activity, stability, and electrochemically active surface area for a variety of OER electrocatalysts in acidic (top, 1 M $H_2SO_4$) and alkaline (bottom, 1 M NaOH) conditions. The x-axis depicts the overpotential at 10 mA cm$^{-2}$ (geometric area) at time t=0, and the y-axis depicts the overpotential at 10 mA cm$^{-2}$ (geometric area) at time t=2 h. A stable catalyst should fall on the diagonal dashed line of the plot, and an active catalyst should be on the bottom left of the plot. The color of each point represents the roughness factor (RF) of the catalyst with light green representing RF = 1, and dark red representing RF > $10^3$. The unshaded white region is the region of interest for screening the catalysts where the overpotentials at 10 mA cm$^{-2}$ (geometric area) at both time t=0 and 2 h are <0.5 V. (Reprinted with permission from ref. 24.. Copyright 2013, American Chemical Society)

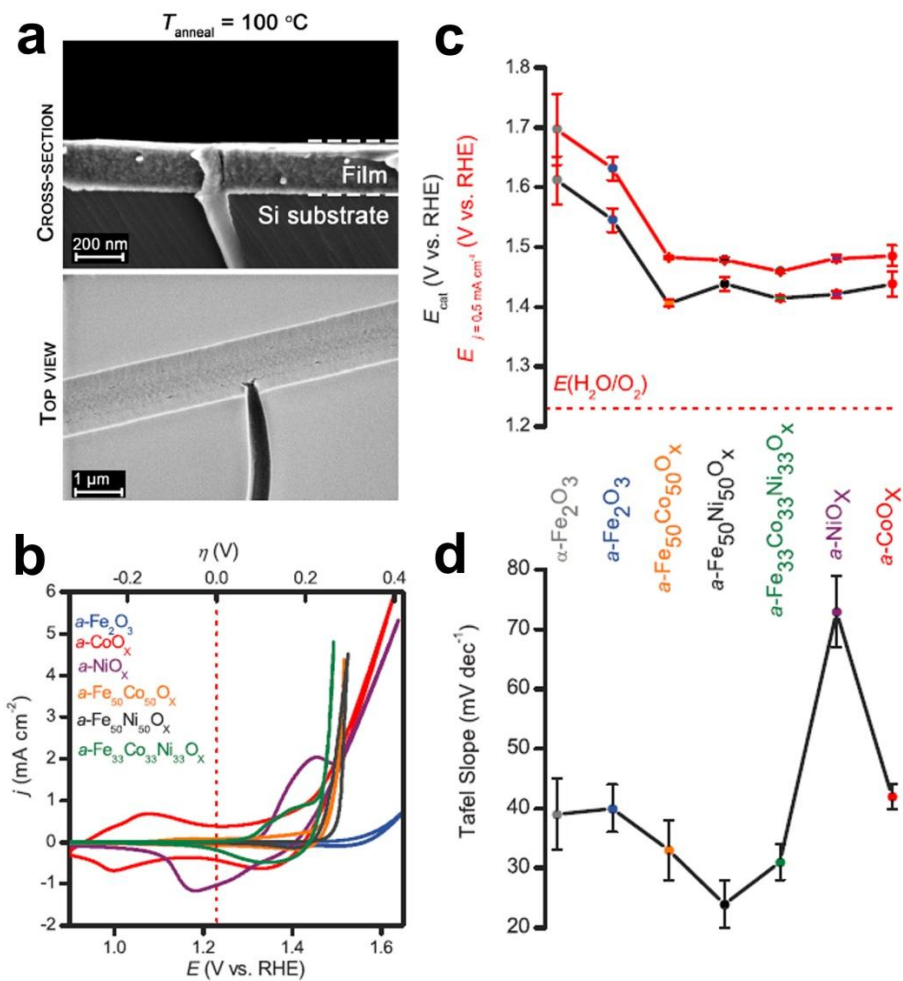

Figure 3. a) cross-section and top-down SEM image of the amorphous $Fe_2O_3$ film by PMOD by annealing at 100°C in air for 1 hour. b) cyclic voltammograms at a scan rate of 10 mV/s, c) overpotentials at 0.5 mV/cm$^2$ (geometric area), d) Tafel slopes of different oxide films prepared by PMOD in 0.1 M KOH. (Reprinted with permission from ref. 67.. Copyright 2013, American Association for the Advancement of Science)

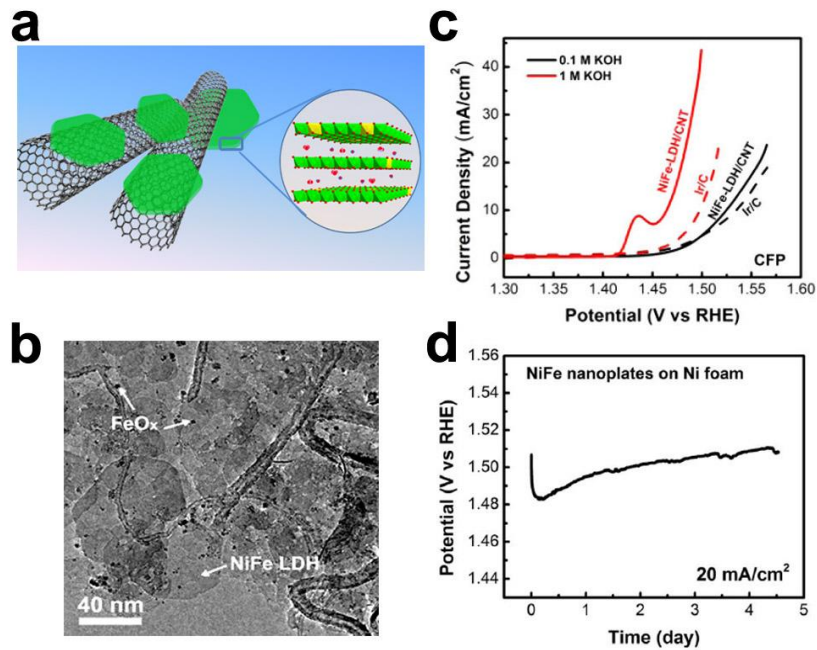

Figure 4. a) schematic view of the hybrid architecture and NiFe LDH structure. b) TEM image of the NiFe-LDH/CNT hybrid. Arrows point to NiFe-LDH plates and iron oxide ($FeO_x$) particles. c) iR-compensated polarization curves of NiFe-LDH/CNT hybrid and commercial Ir/C catalyst on carbon fiber paper (CFP) loaded at 0.25 mg/cm$^2$ for both catalysts electrode in 0.1 and 1 M KOH. d) chronopotentiometry of NiFe LDH nanoplates without CNT loaded onto Ni foam at a loading of ~5 mg/cm$^2$ in 1 M KOH with a constant current density of 20 mA/cm$^2$. (Reprinted with permission from ref. 73.. Copyright 2013, American Chemical Society)

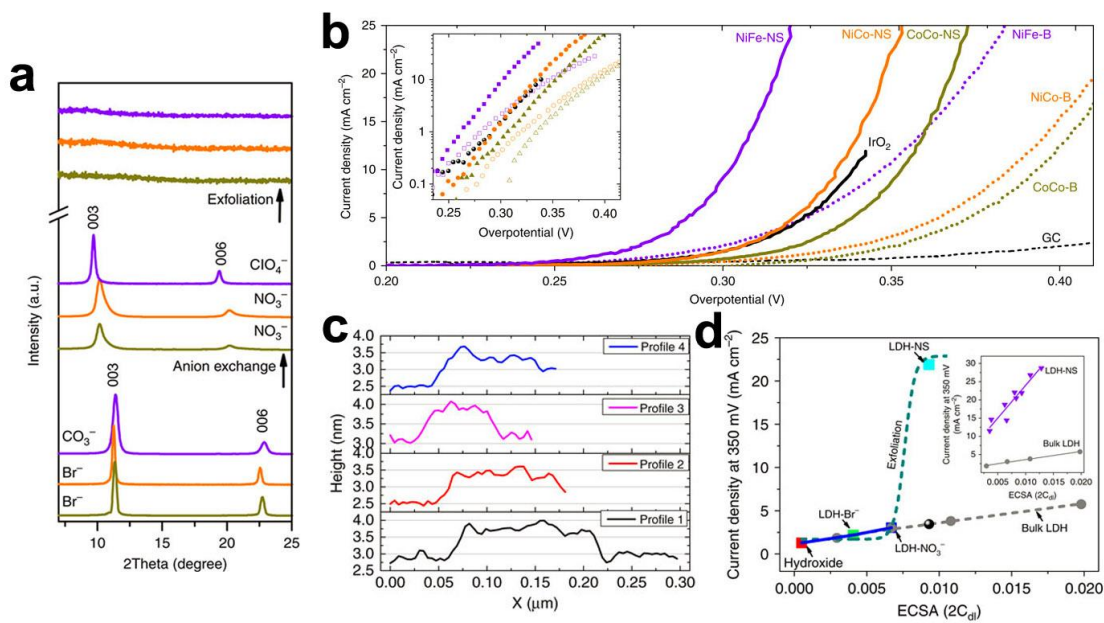

Figure 5. a) XRD spectra of bulk, anion-exchanged and exfoliated LDH (CoCo LDH: yellow-green; NiCo LDH: orange; NiFe LDH: purple). b) polarization curves of bulk and exfoliated LDH at a loading of 0.07 mg/cm$^2$ in 1 M KOH. IrO$_2$ nanoparticles were test side-by-side for comparison. The inset shows the Tafel plots. c) Height profile of the exfoliated NiCo LDH derived from atomic force microscopy (AFM). The average thickness is ~0.8 nm. d) current density at an overpotential of 350 mV plotted versus ECSA (2C$_{dl}$) for different NiCo-based materials. (Reprinted with permission from ref. 78.. Copyright 2014, Rights Managed by Nature Publishing Group)

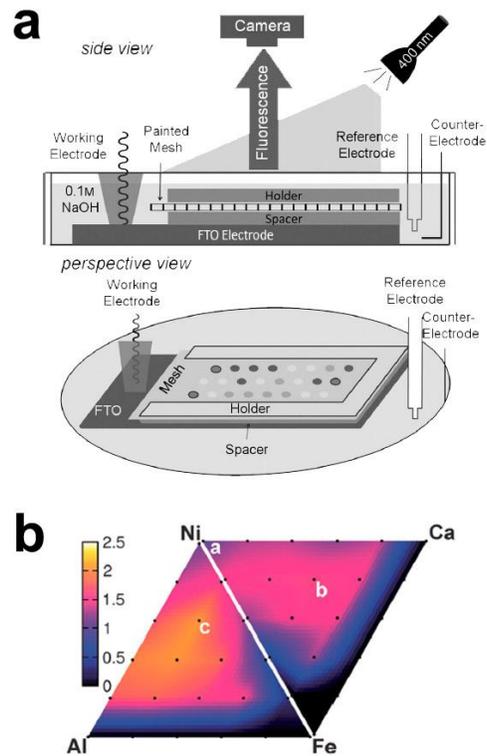

Figure 6. a) schematic diagram of the high throughput electrochemical screening apparatus. b) activity plot with respect to different composition for the triads Ni/Al/Fe and Ni/Ca/Fe. (Reprinted with permission from ref. 79.. Copyright 2012, Wiley-VCH Verlag GmbH & Co. KGaA, Weinheim)

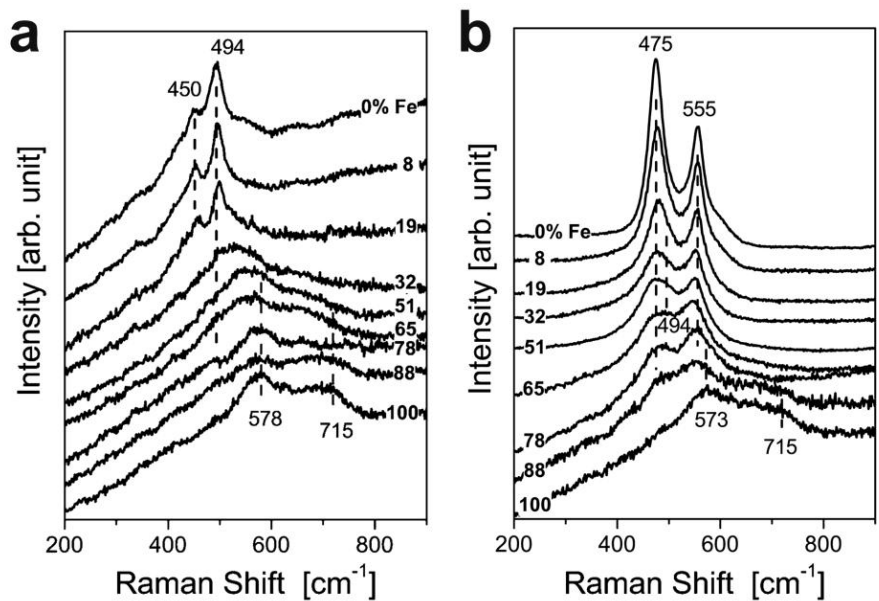

Figure 7. In situ Raman spectra of Ni–Fe catalysts with different composition in 0.1 M KOH at a potential of a) 0.2 V vs Hg/HgO (1 M KOH) and b) 0.6 V vs Hg/HgO (1 M KOH). (Reprinted with permission from ref. 53.. Copyright 2013, American Chemical Society)

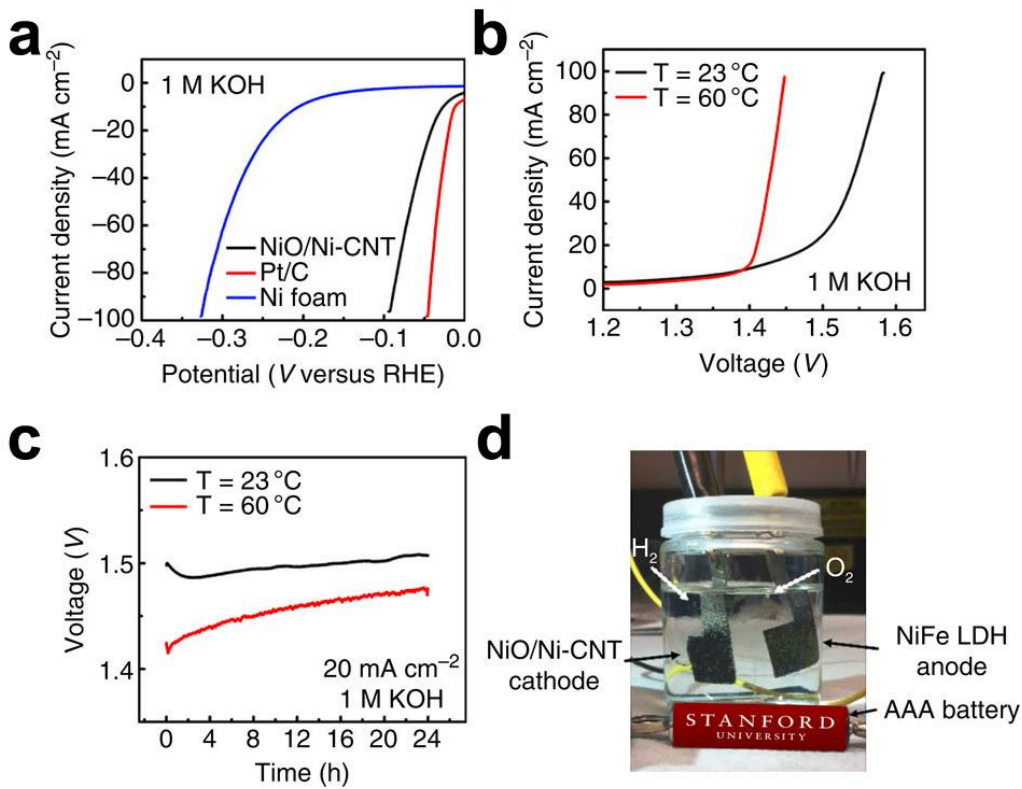

Figure 8. a) Linear sweep voltametry of NiO/Ni-CNT, Pt/C deposited on Ni foam (under the loading of 8 mg cm$^{-2}$) and pure Ni foam at a scan rate of 1 mV/s in 1 M KOH. b) Linear sweepment voltametry of water electrolysis using NiO/Ni-CNT as cathode and NiFe LDH as anode (both deposited on Ni foam under the loading of 8 mg cm$^{-2}$) in 1 M KOH under different temperature. (c) Chonopotentiometry of water electrolysis with NiO/Ni-CNT as cathode and NiFe LDH as anode at a constant current density of 20 mA cm$^{-2}$ in 1 M KOH under different temperature. (d) Demonstration of water electrolyzer powered by an AAA battery with a voltage of 1.5 V. (Reprinted with permission from ref. 12.. Copyright 2014, Rights Managed by Nature Publishing Group)

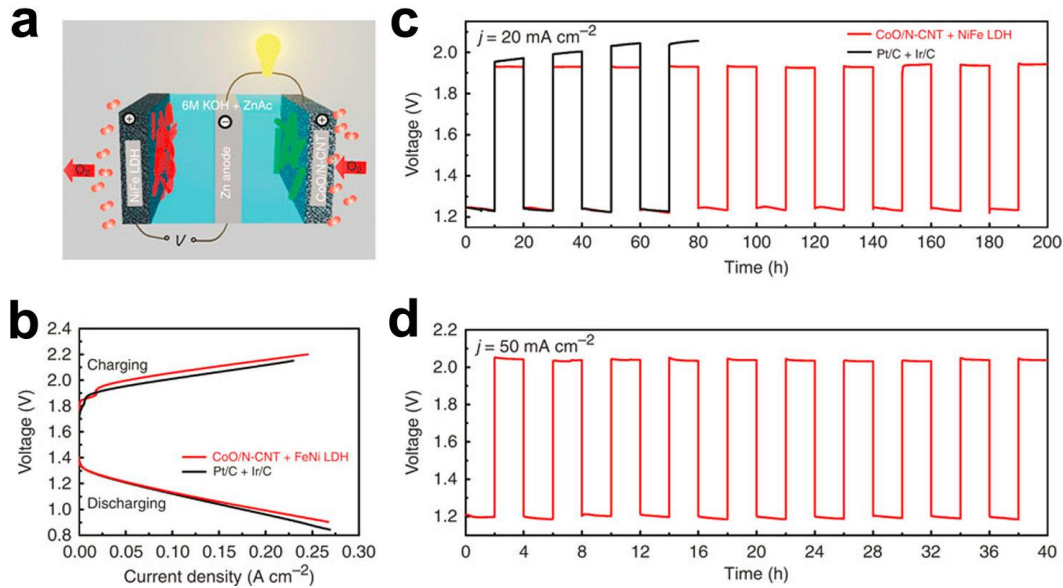

Figure 9. a) Schematic view of the tri-electrode configuration with CoO/NCNT ORR catalyst loaded on CFP electrode (1 mg cm$^{-2}$) and NiFe LDH OER catalyst loaded on Ni foam (~5 mg cm$^{-2}$) electrode for discharge and charge, respectively. (b) Charge/ discharge polarization curves of the rechargeable Zn-air battery using CoO/NCNT and NiFe LDH (red) compared with the one using commercial Pt/C and Ir/C (black). (c) Charge/discharge cycling performance of the rechargeable Zn-air battery using CoO/NCNT and NiFe LDH at 20 mA cm$^{-2}$ with charge/discharge time of 10 h compared with the one using Pt/C and Ir/C. (d) Charge/discharge cycling performance of the rechargeable Zn-air battery using CoO/NCNT and NiFe LDH at 50 mA cm$^{-2}$ with charge/discharge time of 2 h. (Reprinted with permission from ref. 92.. Copyright 2013, Rights Managed by Nature Publishing Group)